\documentclass[aps,reprint,onecolumn,notitlepage]{revtex4-1}
\usepackage{amsmath, amssymb, color, enumitem, graphicx}

\usepackage{natbib}
\usepackage{amsmath,amsbsy,amssymb,amsfonts}

\usepackage{color}
\usepackage{bm}
\usepackage{graphicx}
\usepackage{xfrac,nicefrac}
\usepackage{url}
\usepackage{cleveref}
\usepackage{epstopdf, epsfig}

\renewcommand{\vector}[1]{\mathbf{#1}}

\newcommand{\vv}{\vector{v}}

\newcommand{\vq}{\vector{q}}
\newcommand{\nn}{ \hat{\bf{n}} }
\newcommand{\parD}[2]{\frac{\partial #1}{\partial #2}}
\newcommand{\ordD}[2]{\frac{d #1}{d #2}}

\newcommand{\parDD}[2]{\frac{\partial^2 #1}{\partial #2 ^2}}

\newif\ifshowcomments
\showcommentstrue

\ifshowcomments
\newcommand{\Eastham}[1]{{\color{red}{\em #1}}}
\newcommand{\Moore}[1]{{\color{blue}{\em #1}}}
\newcommand{\Cogan}[1]{{\color{green}{\em #1}}}
\newcommand{\Steinbock}[1]{{\color{cyan}{\em #1}}}
\newcommand{\Wang}[1]{{\color{magenta}{\em #1}}}
\else
\newcommand{\Eastham}[1]{}
\newcommand{\Moore}[1]{}
\newcommand{\Cogan}[1]{}
\newcommand{\Steinbock}[1]{}
\newcommand{\Wang}[1]{}
\fi

\usepackage{soul}

\begin{document}
\title{Multiphase modeling of precipitation-induced membrane formation}
\author{P. S. Eastham}
\email[Email address for correspondence: ]{peastham@math.fsu.edu}
\affiliation{Department of Mathematics, Florida State University,
Tallahassee, FL 32304, USA}
\author{M. N. J. Moore}
\affiliation{Department of Mathematics, Florida State University,
Tallahassee, FL 32304, USA}
\author{N. G. Cogan}
\affiliation{Department of Mathematics, Florida State University,
Tallahassee, FL 32304, USA}
\author{Q. Wang}
\affiliation{Department of Chemistry \& Biochemistry, Florida State University, Tallahassee, FL 32306, USA}
\author{O. Steinbock}
\affiliation{Department of Chemistry \& Biochemistry, Florida State University, Tallahassee, FL 32306, USA}



\begin{abstract}
We formulate a model for the dynamic growth of a membrane developing in a flow as the result of a precipitation reaction, a situation inspired by recent microfluidic experiments. The precipitating solid introduces additional forces on the fluid and eventually forms a membrane that is fixed in the flow due to adhesion with a substrate. A key challenge is that the location of the immobile membrane is unknown \emph{a priori}. To model this situation, we use a multiphase framework with fluid and membrane phases; the aqueous chemicals exist as scalar fields that react within the fluid to induce phase change. To verify that the model exhibits desired fluid-structure behaviors, we make a few simplifying assumptions to obtain a reduced form of the equations that is amenable to exact solution. This analysis demonstrates no-slip behavior on the developing membrane without \emph{a priori} assumptions on its location. The model has applications towards precipitate reactions where the precipitate greatly affects the surrounding flow, a situation appearing in many laboratory and geophysical contexts including the hydrothermal vent theory for the origin of life. More generally, this model can be used to address fluid-structure interaction problems that feature the dynamic generation of structures.
\end{abstract}

\maketitle

\section{Introduction}
One hypothesis for the ``origin of life" is that the first biomolecules were formed in undersea hydrothermal vents. In this theory, passive, anisotropic diffusion across a  membrane supports the transmembrane gradients necessary for the first biochemical molecules \citep{martin2008hydrothermal}. An experimental approach to study this theory examines simpler systems in microfluidic reactors which allow for the controlled study of the prebiotic chemistry in hydrothermal vent chimneys \citep{Barge17}. 

Microfluidic devices have become an important tool in modern chemistry and biomedical analytics \citep{Nge23}. One application is the possibility of a ``lab on a chip", i.e.~the miniaturization of chemical separation and analysis procedures onto a disposable device as small as a few square centimeters. The devices are typically made from etched glass or lithographically-processed elastomers and the fluid flow is usually controlled mechanically by external pumps or electrically via electro-osmotic flows \citep{Mark10}.

Recent studies have used microfluidic methods to form inorganic membranes within Y-shaped devices \citep{batista2015growing, Wang17}. The membranes result from chemical reactions between two different solutions that are injected separately but later merge in a long reaction channel that brings the reactants into direct contact. This merging is usually performed under low Reynolds number (Re)  conditions and for miscible liquids, such as aqueous solutions of NaOH, and a dissolved metal salt, such as NiCl$_2$. At the reactive interface between these liquids, a precipitate, such as Ni(OH)$_2$, swiftly forms a thin porous membrane (Figure 1). This precipitate reaction typically involves the formation of microscopically small colloid particles and their aggregation or addition to the membrane. This phenomenon is related to so-called ``chemical gardens" which consist of thin cylindrical precipitate membranes separating a metal salt solution from silicate or hydroxide solutions \citep{roszol2011controlling,makki2009hollow}.

\begin{figure}[h]
\centering
\includegraphics[scale=1]{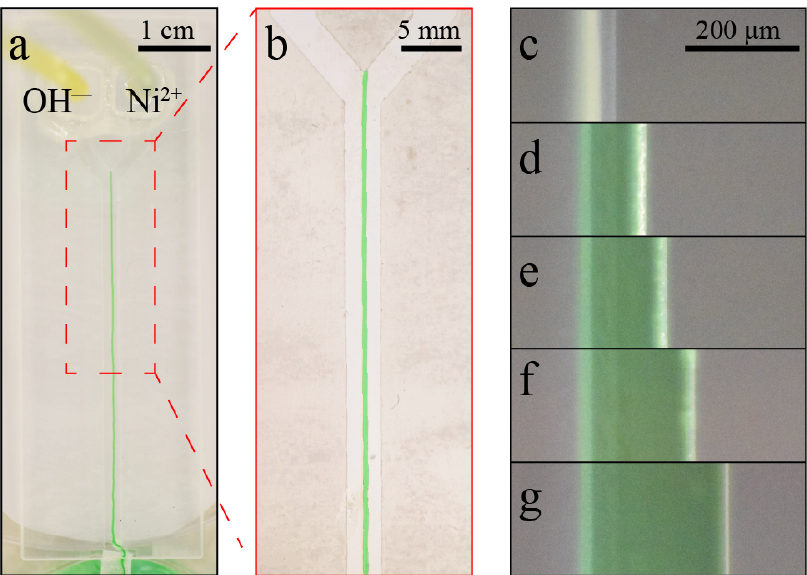}
\caption{Inorganic precipitate membrane formed in a microfluidic channel. (a) The photograph of the resulting membrane after 2 h with 0.5 M NaOH and 0.5 M NiCl$_2$ solutions being injected simultaneously into the microfluidic device. The mixing part of the channel is 50 mm long, 2 mm wide, and approximately 130 $\mu$m high. (b) A magnified view of a selected area from (a).  (c-g) A sequence of micrographs showing the unidirectional thickening process after (c) 1 min, (d) 15 min, (e) 30 min, (f) 60 min, and (g) 120 min. Scale bars correspond to (a) 1 cm, (b) 5 mm, and (c) 200 $\mu $m. }
\label{fig:experiment}
\end{figure}

\citet{ding2016wavy} showed experimentally that membrane thickness increases with the square root of time, indicating diffusion-controlled growth. The membrane thickening occurs only in the direction of the metal salt solution (e.g.~NiCl$_2$) and not in the direction of the anionic precipitation partner (e.g.~OH$^-$), indicating that the membrane is more permeable to anions than cations. This phenomenon has been qualitatively explained by the charged nature of the membrane that suppresses the transmembrane transport of the positive metal ion \citep{batista2015growing, Wang17}. 

The modeling challenges presented by this experiment involve a confluence of topics that have been studied before, namely ionic reactions \citep{sircar2013effect,yang2019quasi}, precipitation \citep{zhang2010mathematical, agarwal2014solute}, passive diffusion through a membrane \citep{mouritsen2005life, cogan2009incorporating, ding2016wavy}, and fluid-structure interaction \citep{vogel1996life, childress2012fluid, ristroph2012sculpting, strychalski2015poroelastic,strychalski2016intracellular, moore2017riemann, quaife2018boundary}. The particular combination of these aspects provides the opportunity for a new model that captures them all. One key challenge is that the “structure” in this problem is generated dynamically according to equations governing the chemistry. We choose to model the fluid-structure combination as a single multiphase material: one component “fluid” or solvent and one component “structure” or precipitate membrane. Such multiphase models have proven useful in a variety of complex-fluid applications, such as bacterial biofilms \citep{cogan2004role,cogan2010multiphase}, tumor-growth \citep{byrne2003modelling, preziosi2009multiphase, frieboes2010diffusional, sorribes2019biomechanical}, and biological membranes \citep{magi2017modelling}; their formulation is based on averaging momentum and stresses in separated, multi-component fluids \citep{drew1983mathematical, drew2006theory}.


The multiphase framework developed here builds on previous ones \citep{cogan2010multiphase, zhang2010mathematical, leiderman2011grow}, but with some keys differences that are guided by a combination of physical principles, model simplicity, and the micro-fluidic experiments mentioned above. First, our formulation conserves the total mass of the components — solvent, dissolved species, and precipitate membrane — throughout evolution. In particular, the model accounts for changes in solute concentrations that result from the formation of new membrane and the associated exclusion of solvent volume. This effect is neglected in previous models that treat reaction chemicals as scalar fields distinct from the multiphase material, but is essential for overall mass conservation. The treatment of reaction chemicals as additional components of a multiphase material has been successfully modeled by many \citep[see]{Nunziato1980,yang2019quasi} but greatly complicates the analysis, interpretation, and simulation of the governing equations. Second, by making certain choices in the averaging procedure for the multicomponent stress, our formulation becomes equivalent to an incompressible Brinkman system with variable permeability. This equivalence is important for a few reasons. First, it guarantees that the model reduces to the Stokes equations in the fluid limit and to Darcy’s equation in the porous-medium limit \citep{brinkman1949calculation, durlofsky1987analysis, hill2001first}. In particular, it guarantees that when membrane is fully developed, the interface behaves as an impermeable surface with a no-slip condition on fluid velocity. As demonstrated in section \ref{sec:poiseuille}, many existing multiphase models fail to exhibit this behavior \citep{breward2003multiphase,cogan2005channel,cogan2010multiphase, sorribes2019biomechanical}, as they were developed primarily for highly permeable systems. Second, the equivalence to Brinkman significantly simplifies the overall structure of the partial differential equation (PDE) system by eliminating certain cross-terms in the stress divergence that arise in other models. This simplification is one key that will allow reduction to a non-trivial case where chemical and phase dynamics can be solved exactly.

To demonstrate that the new framework possesses the desired properties listed above, we consider a simplified system in which the incoming reactant concentrations are held fixed via chemostat \citep{rubinow1975introduction}. By assuming parallel flow and neglecting solute diffusion, the governing equations reduce to a planar system of ordinary differential equations (ODEs). This nonlinear system can be linearized around the fixed points, and eigenvalue analysis provides an estimate for the rate at which new membrane forms. Moreover, we find that the equation for the aqueous product is a second-order nonlinear ODE known as the Ricatti equation \citep{ricatti1724animadversiones,tenenbaum1985ordinary}. Exact solutions to the Ricatti equation give explicit formulas for the time dependence of the chemical product and, consequently, the formation of new membrane. Once the membrane dynamics are known exactly, the flow profile can be obtained through the numerical solution of a simple boundary value problem (BVP). Access to the resulting flow profile allows a careful comparison between variants of the multiphase framework. In particular, we demonstrate that the framework developed here properly captures the transition from one-channel to two-channel flow as membrane develops.

The paper proceeds as follows: in section \ref{sec:model} we develop the governing equations for both the reacting chemicals and multiphase material such that the total mass is conserved. Section \ref{sec:analysis} contains analysis and results based on simplifying assumptions. These assumptions generate a reduced form for an idealized scenario which can be solved with a combination of analytic and simple numerical methods. Finally in section \ref{sec:discussion} the predictive power of the model and further applications are discussed.

\section{Mathematical Model} \label{sec:model}
The model requires the accurate description of several aspects of the experiment, including the flow transport of the two ionic species and their reaction to form a product, the precipitation of the product out of solution, and finally the response of the bulk fluid motion to the dynamically-generated precipitate membrane. Advection-Diffusion-Reaction (ADR) equations are derived for the aqueous chemical concentrations, while the fluid and membrane dynamics are described by multiphase mass and momentum balance equations. In many multiphase models, either constituent can be viscous, viscoelastic, poroelastic, or otherwise. Here, since the membrane adheres to the substrate, it can be treated as an immobile solid, leading to considerable simplifications. 

We assume that aqueous reactants and product contribute mass, but not volume, to the fluid phase. The solvent and membrane each have their own distinct mass densities, and any arbitrary control volume can be divided into solvent and membrane volume fractions. The formation of new membrane involves the precipitation of product out of solution and the sequestration of  solvent. A key modeling assumption is that the volume of fluid sequestered equals the volume of the resulting membrane. As shown in section \ref{sec:massbalanceeqns}, this assumption ensures incompressibility of the phase-averaged velocity field, i.e.~the so-called Darcy velocity.

We now detail the model equations. First we derive evolution equations for the reaction of aqueous ionic species, then we list mass balances for all chemical species as well as solvent and membrane phases, and finally we describe the momentum equation for the fluid. In the end we obtain a closed, coupled PDE system governing the chemistry and physics of the system, where total mass is conserved throughout aqueous reactions and phase transitions.

\subsection{Model for Chemical Reactions} \label{sec:modelchemicals}
In this section we derive equations for the chemical reactions. We follow the ``nucleation and growth" model of precipitation \citep{matsue2018role} and separate the reaction into two sequential parts: in the first, two reactants come together to form an aqueous product, and the second describes the aggregation of the aqueous product into a solid precipitate. While many aqueous chemical reactions do not alter the solution volume significantly, the formation of a membrane excludes fluid volume and therefore can alter the local concentration of the dissolved species. Accordingly, our model neglects the volume occupied by the aqueous species but does account for changes in species concentration that are due to the precipitated solid excluding fluid volume. This effect introduces additional terms in the aqueous reaction equations that are required for mass conservation. To our knowledge these additional terms are not accounted for in the multiphase precipitation literature that treats aqueous chemicals as scalar fields distinct from of the multiphase material.

The aqueous reaction is written as a generic net ionic equation 
\begin{equation}
aA\text{(aq)} \,+\,bB\text{(aq)} \to cC\text{(aq)} \label{eq:aqueous}
\end{equation}
where $A\text{(aq)}$, $B\text{(aq)}$, and $C\text{(aq)}$ are chemicals in the aqueous phase and $a$, $b$, and $c$ are their respective stoichiometric coefficients; the precipitation reaction is written simply as 
\begin{equation}
    C\text{(aq)} \to C\text{(s)}\,.\label{eq:precipreaction}
\end{equation}
As a concrete example consider the reaction described in the introduction, 
\begin{gather}
\text{Ni}^{2+}\text{(aq)} + 2\text{(OH)}^-\text{(aq)} \to \text{Ni(OH)}_2\text{(aq)} \\[1em]
\text{Ni(OH)}_2\text{(aq)} \to \text{Ni(OH)}_2\text{(s)}\,. 
\end{gather} 
Then $A=\text{Ni}^{2+}$, $B=\text{(OH)}^-$ and $C=\text{Ni(OH)}_2$ and $a=c=1$, $b=2$. 

The aqueous chemicals will be measured with a variable for the \emph{number} of chemicals per unit \emph{solvent} volume, i.e.~molarity, which we will call $\psi_i$ for chemical species \mbox{$i\in\{A,B,C\}$}. Reaction rates depend on a reactants' molarity, and molarity can change due to two independent factors: either the number of molecules changes due to the aqueous reaction, or the solvent volume changes due to precipitation. Because either one can occur in a precipitation reaction, these two competing effects must be carefully considered when formulating the reaction equations.

\begin{figure}[h]
    \begin{center}
	\includegraphics[width= 0.9 \linewidth]{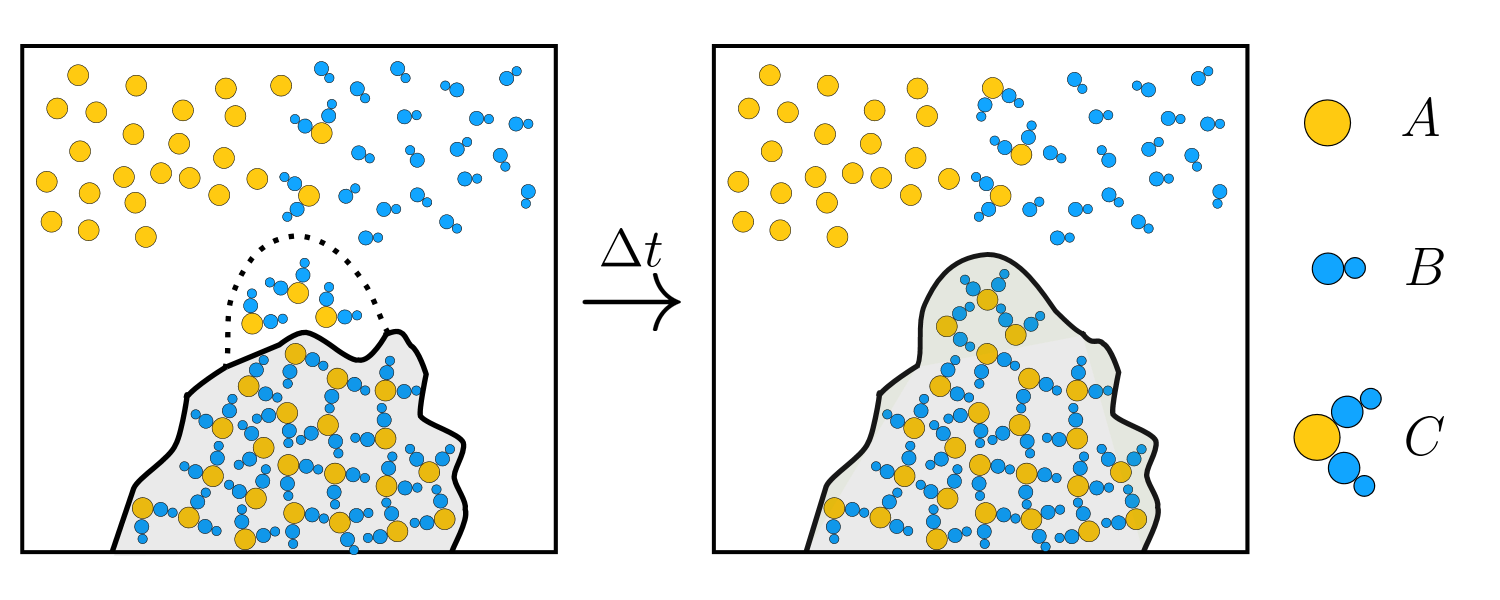}
	\end{center}
	\caption{Schematic of precipitate reaction in control volume. Precipitation causes solution (white) to transform into membrane (shaded) after a certain concentration threshold is reached of aqueous product $C$. Aqueous chemicals $A$, $B$ and $C$ are volumeless scalar fields while the solvent and membrane is treated as a multiphase material. The volume of membrane gained is exactly equal to the volume of solvent lost.}
	\label{fig:contmech}
\end{figure}

We begin by deriving equations for how the aqueous reaction proceeds in a spatially homogeneous environment; later in section \ref{sec:massbalanceeqns} the effects of advective and diffusive spatial fluxes will be added. Suppose the chemicals exist in some aqueous solution of fixed control volume $V_0$. The chemicals undergo both the aqueous and precipitation reactions which results in fluid mass and solvent volume being converted to membrane mass and volume (see figure \ref{fig:contmech}). The fluid component has mass 
\begin{equation}
\mathcal{M}_f = \rho_f\theta_sV_0 = \left(\rho_s + \sum M_i\psi_i\right)\theta_sV_0 \label{eq:fluidmass}
\end{equation} 
where $\rho_s$ is the constant solvent mass density (without any reactants or products present), $\theta_s$ is the solvent volume fraction, and $M_i$ is the molar mass of chemical species $i\in\{A,B,C \}$. The summation represents the contribution of the chemical species to fluid mass, so that the {\em fluid} mass density $\rho_f$ is not constant.
The membrane component has mass $\mathcal{M}_m = \rho_m\theta_m V_0$ where $\rho_m$ is the constant membrane mass density and $\theta_m$ is the membrane volume fraction. Physically, membrane mass is composed of both precipitated chemical $C$ and sequestered solvent mass.

The change in $\psi_i$ \emph{purely due to aqueous reaction}, i.e.~no precipitation, can be modeled as a second-order kinetics reaction 
\begin{equation}
\dot{\psi}^{(aq)}_A = -ar\psi_A\psi_B,\qquad \dot{\psi}^{(aq)}_B = -br\psi_A\psi_B,\qquad \dot{\psi}^{(aq)}_C = cr\psi_A\psi_B\,. \label{eq:aqreact}
\end{equation}
where the dot indicates a derivative with respect to time, and $r$ is the rate of aqueous reaction per chemical concentration. More general power laws are sometimes used to model chemical kinetics, but here we use purely second-order kinetics for simplicity \citep[see][pp. 573 - 575]{changchemistry2013}. None of the analysis, however, depends specifically on this choice and the results could be carried forward for other kinetics.

To derive equations for the change in $\psi_i$ \emph{purely due to precipitation} we appeal to ideas from continuum mechanics. The concentration of ions $A$ in the control volume is written as $\psi_A=n_A/(\theta_sV_0)$ where $n_A$ is the number of $A$ ions in $V_0$. Note that this formulation makes explicit the dependence of $\psi_A$ on both $n_A$ and $\theta_s$. Consider the change in a small increment of time $\Delta t$. Then the time-dependent variables are updated so that
\begin{equation}
    \psi_A + \Delta\psi_A = \frac{n_A}{(\theta_s+\Delta\theta_s)V_0}\,.
\end{equation}
Recall that $n_A$ is constant during precipitation as only $C\text{(aq)}$ precipitates. Approximating for small $\Delta \theta_s$ and neglecting higher-order terms gives
\begin{equation}
    \psi_A + \Delta\psi_A = \frac{n_A}{\theta_sV_0}\left(1 - \frac{\Delta\theta_s}{\theta_s}\right) = \psi_A\left(1 - \frac{\Delta\theta_s}{\theta_s}\right) \, .
\end{equation}
Then, cancelling the $\psi_A$, dividing both sides by $\Delta t$, and letting $\Delta t \to 0$ gives the change in $\psi_A$ purely due to precipitate reaction as $\dot{\psi}^{(p)}_A=-\psi_A\dot{\theta}_s/\theta_s$. By symmetry, a similar formula holds for $\dot{\psi}^{(p)}_B$.
Note that both of these are essentially applications of the product rule for $\partial_t(\psi_i\theta_s)=0$, which physically means that the total number of ions of $i\in\{A,B\}$ in the control volume does not change in time \emph{due to precipitation}.

A similar procedure can be followed for $\psi_C$, except now the number of aqueous chemicals $n_C$ changes as $C\text{(aq)}$ precipitates into membrane,
\begin{equation}
    \psi_C+\Delta\psi_C = \frac{n_C+\Delta n_C}{(\theta_s+\Delta\theta_s)V_0} \, .
\end{equation} 
Above, both $n_C$ and $\theta_s$ change in time. Expanding both expressions while linearizing for small $\Delta\theta_s$, dividing by $\Delta t$, and taking the limit as $\Delta t\to 0$ one obtains \mbox{$\dot{\psi}^{(p)}_C = - \psi_C\dot{\theta}_s/\theta_s + \dot{n}_C/\theta_s$}. The first term in this expression is analogous to those obtained for reactants $A$ and $B$, and simply describes the effect on concentration when solvent volume is changing. The second term, however, is new and describes the effect on $\psi_C$ as aqueous $C$ molecules are converted into membrane. We write $\dot{n}_C = \alpha\dot{\theta}_s$ where the specific value of $\alpha$ will be found shortly to guarantee conservation of mass throughout the entire reaction. The expressions for the rate of change of aqueous chemical concentrations due to precipitation are thus:
\begin{equation}
\dot{\psi}^{(p)}_A = -\psi_A\dot{\theta}_s/\theta_s,\qquad \dot{\psi}^{(p)}_B = -\psi_B\dot{\theta}_s/\theta_s, \qquad \dot{\psi}^{(p)}_C = -\psi_C\dot{\theta}_s/\theta_s + \alpha\dot{\theta}_s/\theta_s\,. \label{eq:chemprecip}
\end{equation}

Assuming that the aqueous and precipitate reactions act independently, $\dot{\psi}_i = \dot{\psi}^{(aq)}_i + \dot{\psi}^{(p)}_i$, gives
\begin{subequations}\label{eq:chemreacthomo}
\begin{align}
\dot{\psi}_A & = -ar\psi_A\psi_B -\psi_A\dot{\theta}_s/\theta_s \label{eq:homA} \\
\dot{\psi}_B & = -br\psi_A\psi_B -\psi_B\dot{\theta}_s/\theta_s \label{eq:homB} \\
\dot{\psi}_C & = cr\psi_A\psi_B -\psi_C\dot{\theta}_s/\theta_s + \alpha \dot{\theta}_s/\theta_s \label{eq:homC}
\end{align}
\end{subequations}
These equations describe the dynamics of aqueous species concentrations in the absence of spatial fluxes.

To obtain the value of $\alpha$ that guarantees conservation of mass, we again apply a continuum mechanics argument. The change in membrane mass after a small time step is $\Delta \mathcal{M}_m = \rho_m\Delta \theta_m V_0$. To simplify the expression for change in fluid mass, we expand $\Delta \mathcal{M}_f$ while neglecting second order terms to get
\begin{equation}
    \Delta \mathcal{M}_f = V_0\rho_s\Delta \theta_s + V_0\theta_s\sum_i M_i\Delta\psi_i + V_0\Delta \theta_s\sum_i M_i\psi_i \, .
\end{equation}
Replacing the $\Delta \psi_i$ with their respective differential terms in equations \eqref{eq:chemreacthomo} and performing some algebraic manipulation produces
\begin{equation}
    \Delta \mathcal{M}_f = V_0\rho_s\Delta\theta_s + V_0\theta_s\psi_A\psi_B(cM_c - aM_A - bM_B) + V_0M_C\Delta n_C \, .
    \label{DeltaMf}
\end{equation}
Conservation of mass during the aqueous reaction \eqref{eq:aqueous} implies 
\begin{equation}
aM_A+bM_B=cM_C\, .
\label{eq:aq_mass_cons}
\end{equation}
Thus, the term in parenthesis in \eqref{DeltaMf} vanishes. Meanwhile, conservation of mass of the entire system implies $\Delta \mathcal{M}_m = -\Delta \mathcal{M}_f$, i.e.~the mass lost by the fluid equals the mass gained by membrane. Additionally, the assumption that fluid volume is converted perfectly to membrane volume implies $\Delta \theta_m = -\Delta \theta_s$. Using the respective definitions of $\mathcal{M}_i$ and solving for $\Delta n_C$ gives
\begin{equation}
    \Delta n_C = \left(\frac{\rho_m-\rho_s}{M_C}\right)\Delta \theta_s \, .
\end{equation}
Diving by $\Delta t$ and taking the limit $\Delta t\to 0$ gives $\dot{n}_C = \alpha\dot{\theta}_s$ where $\alpha = (\rho_m - \rho_s)/M_C$. Physically, this value of $\alpha$ corresponds to the concentration of $C\text{(aq)}$ that must leave the fluid phase during precipitation in order for mass to be conserved.

The reaction equations derived in this section, along with the specific $\alpha$ term, will be used to provide reaction terms for the chemistry mass balance equations, as described in the next section.

\subsection{Mass Balance Equations} \label{sec:massbalanceeqns}

In the experiments, the aqueous reaction occurs within the flow of a microfluidic device and therefore spatial fluxes must be considered. To describe these fluxes, consider the general conservation law for the \emph{chemical mass per unit control volume} $\phi=M_i\psi_i\theta_s$,
\begin{equation}
\parD{\phi}{t} + \nabla\cdot\vector{J} = \Gamma 
\end{equation}
where $\vector{J}$ is the flux of $\phi$ and $\Gamma$ is a transfer term for the rate that $\phi$ enters the system. 

We choose $\vector{J}$ to account for advection and diffusion of the chemical concentrations,
\begin{align}
\parD{(M_A\psi_A\theta_s)}{t} + \nabla\cdot 
\big(M_A\psi_A\theta_s\vv_s - \kappa_A\nabla(M_A\psi_A)\big) & = \Gamma_A \label{eq:inhomA} \\
\parD{(M_B\psi_B\theta_s)}{t} + \nabla\cdot\big(M_B\psi_B\theta_s\vv_s - \kappa_B\nabla(M_B\psi_B)\big) & = \Gamma_B \label{eq:inhomB}\\
\parD{(M_C\psi_C\theta_s)}{t} + \nabla\cdot\big(M_C\psi_C\theta_s\vv_s - \kappa_C\nabla(M_C\psi_C)\big) & = \Gamma_C \label{eq:inhomC}
\end{align}
where $\vv_s$ is the (tracer) velocity of the solvent and $\kappa_i$ are diffusion coefficients which possibly depend on the solvent volume fraction. 
Note that the diffusive flux used above transports mass according to gradients in {\em molarity} $\psi_i$, not gradients in $\phi_i$. This choice produces the physically realistic steady state of uniform molarity in a quiescent, non-reacting fluid that has inhomogeneous volume fraction.

Assuming that the reactions and spatial fluxes act independently, the $\Gamma_i$ correspond to the rates given in equations \eqref{eq:chemreacthomo}. Rearranging and multiplying each equation by its respective molar mass $M_i$ gives
\begin{subequations}
\label{Gammas}
\begin{gather}
    \Gamma_A = -arM_A\theta_s\psi_A\psi_B \\
    \Gamma_B = -brM_B\theta_s\psi_A\psi_B \\
    \Gamma_C = crM_C\theta_s\psi_A\psi_B + \alpha M_C\dot{\theta}_s
\end{gather}
\end{subequations}
where $\alpha=(\rho_m-\rho_s)/M_C$. Now that mass balance equations for the chemistry are established, mass balance equations for the multiphase solvent-membrane system are needed. 

A simple but necessary assumption is that our volume is occupied by only solvent and membrane, i.e.~there are no ``voids". This no-void assumption implies
\begin{equation}
    \theta_s  + \theta_m = 1 \label{eq:consvolume}
\end{equation}
everywhere. Mass balances for the solvent and membrane phases provide
\begin{align} 
& \parD{(\rho_s\theta_s)}{t} + \nabla\cdot(\rho_s\theta_s\vv_s)= R_s  \label{eq:pre-massbalanceS} \\
& \parD{(\rho_m\theta_m)}{t}  = R_m \label{eq:pre-massbalanceM}
\end{align}
where $R_i$ denotes the rate of mass added to phase $i$. Equation \eqref{eq:pre-massbalanceM} has no advective term since the membrane is assumed to be immobile. 

To ensure conservation of total mass, the rates $R_m$ and $R_s$ must be related. To derive this relationship, let $V_0$ be an arbitrary control volume. The total mass (of all components) in $V_0$ is
\begin{equation}
\mathcal{M}(V_0) = \int_{V_0} {\rho_s\theta_s + \rho_m\theta_m + \sum M_i\psi_i\theta_s}\,\, dV
\end{equation}
Summing the five mass balance equations, \eqref{eq:inhomA}--\eqref{eq:inhomC} and \eqref{eq:pre-massbalanceS}--\eqref{eq:pre-massbalanceM}, integrating over $V_0$, and applying the divergence theorem gives
\begin{equation}
\label{massb1}
\ordD{}{t}\mathcal{M}(V_0) 
+ \int_{\partial V_0} \underbrace{\left( \rho_s \theta_s \vv_s + \sum \vector{J}_i \right) \cdot \nn }_{\text{boundary flux}} \,\, dS  
= \int_{V_0} \underbrace{R_s + R_m + \sum \Gamma_i}_{\text{transfer \& reaction}}\,\, dV \, . 
\end{equation}
where $\nn$ is the outward unit normal vector. Summing equations \eqref{Gammas} and applying \eqref{eq:aq_mass_cons} gives
$ \sum \Gamma_i = \alpha M_C\dot{\theta}_s$.
For the sake of obtaining a relationship between $R_s$ and $R_m$, briefly consider the case of zero boundary flux. Then to conserve total mass within any control volume, we must have
$R_s + R_m + \alpha M_C\dot{\theta}_s = 0$
holding point-wise. Substituting the value of $\alpha$, using the no-void assumption \eqref{eq:consvolume} and the definition of $R_m$ in \eqref{eq:pre-massbalanceM} gives
\begin{equation}
\label{RsRm}
R_s = -\frac{\rho_s}{\rho_m} \, R_m
\end{equation}
This relationship is required for overall mass conservation. 

Now consider the so-called Darcy velocity field $\vq_s = \theta_s\vv_s$. 
Substituting the value of $R_s$  from equation \eqref{RsRm} into equation \eqref{eq:pre-massbalanceS}, using a consequence of the no-void assumption ($\dot{\theta}_s=-\dot{\theta}_m$) and substituting $R_m$ by its value in equation \eqref{eq:pre-massbalanceS} implies that the Darcy velocity field is in fact incompressible
\begin{equation}
\label{incomp}
\nabla \cdot \vq_s = 0 \, .
\end{equation}
Now return to \eqref{massb1} and consider the entire domain $\Omega$ with total mass $\mathcal{M}=\mathcal{M}(\Omega)$. From the above argument, the right-hand-side of this equation vanishes as a necessary condition on mass conservation. Then applying incompressibility \eqref{incomp}, gives the total mass balance
\begin{equation}
\label{massb3}
\ordD{\mathcal{M}}{t} =
- \int_{\partial \Omega} \sum M_i \big( \vq_s \psi_i - \kappa_i \nabla  \psi_i \big) \cdot \nn  \,\, dS  \, . 
\end{equation}
As expressed in this equation, the total mass of the system is conserved as long as the chemical flux at the boundary vanishes. More generally, the total mass of the system can change according to how much chemical mass is being injected or removed via the boundary flux terms.

We now specify our choice for the form for the precipitation term $R_m$. Although complicated models of precipitation exist \citep{matsue2018role,ostapienko2018mathematical}, we employ a simple model in which the rate of membrane mass growth is proportional to the amount of product, provided that the product concentration exceeds some precipitation threshold, i.e.
\begin{equation}
R_m = \beta \psi_C\theta_s\mathcal{H}(\psi_C - \psi_C^*) \label{eq:preciprate}
\end{equation}
where $\mathcal{H}$ is the standard Heaviside function and $\psi_C^*$ is the concentration threshold for precipitation to occur. 

\subsection{Momentum Balance Equations} \label{sec:momentumbalanceeqns}
Since inertial effects are assumed negligible ($\text{Re}\ll 1$), the solvent momentum balance can be written as
\begin{equation} 
\nabla\cdot\mathbf{T} - \theta_s\nabla P - \xi\vv_s = \mathbf{0} \label{eq:momentumfirst} 
\end{equation}
where $\mathbf{T}$ is the multiphase stress tensor, $P$ is a so-called common pressure that is shared by the fluid and membrane phases \citep{cogan2010multiphase}, and $\xi$ is a friction coefficient. Deviating slightly from the predominant multiphase literature, we chose the form of the stress tensor as 
\begin{equation}
\mathbf{T} = \eta \big(\nabla \vq_s + \nabla \vq_s^\top\big)
\end{equation}
where $\eta$ is the fluid viscosity. In particular, since $\vq_s = \theta_s \vv_s$, we have placed the fluid volume fraction $\theta_s$ {\em inside} the gradient, whereas most multiphase models place the $\theta_s$ outside of the gradient but inside the divergence \citep{byrne2003modelling,cogan2005channel,cogan2010multiphase}. Such a choice must be made for model closure, and neither is fully justified by first principles. We make the above choice to obtain equivalence to the Brinkman system, which offers considerable gains in model tractability.

The incompressibility of the Darcy velocity, while being notable in itself, allows the momentum equation to be transformed into something more familiar. Applying the divergence-free property to equation  \eqref{eq:momentumfirst} reduces it to a Brinkman equation with variable coefficients
\begin{equation}
    \eta \nabla^2 \vq_s - \frac{\xi}{\theta_s} \vq_s = \theta_s \nabla P\,. \label{eq:momentum1}
\end{equation}
We can now see that equation \eqref{eq:momentumfirst}, equivalently equation \eqref{eq:momentum1}, does not have any cross-derivative terms that appear in \citet{cogan2010multiphase,keener2011kinetics}. Because the membrane is assumed immobile ($\vv_m\equiv \vector{0}$), no momentum equation is needed for it.


The friction coefficient $\xi$ should be chosen in such a way that, at high membrane volume fraction, friction becomes the dominant effect in equation \eqref{eq:momentum1}. The choice made here, and mentioned briefly in \citet{leiderman2013influence}, is to use the Kozeny-Carman (KC) formula for permeability as it depends on porosity \citep{dullien2012porous}. In the present notation, the KC relationship gives the friction coefficient as
\begin{equation}
    \xi_{KC}(\theta_s) = h\frac{(1-\theta_s)^2}{\theta_s} 
\end{equation}
where $h$ is an arbitrary constant. This friction coefficient will provide the desired no-slip behavior in the membrane limit $\theta_m\to 1$. \citet{angot1999analysis} discusses the implications of a similar singular friction term in a Brinkman system, although their model does not include the solvent volume fraction term in front of the pressure gradient and only applies to domains with spatially discontinuous volume fractions; the current fluid-membrane model generalizes this notion by being able to account for smooth spatial and temporal gradients of the volume fraction. 

As an alternative to the singular $\xi_{KC}$, the friction coefficient can be chosen to be a (non-singular) Hill function, as used in \citet{leiderman2011grow, leiderman2013influence},
\begin{equation}
    \xi_H(\theta_s) = h\frac{(1-\theta_s)^n}{K^n + (1-\theta_s)^n}\,.
\end{equation}
The use of Hill functions is largely empirical, although it has significant advantages in that it is finite in the membrane limit and therefore more numerically stable. Additionally, $K$ determines the half-saturation point and $n$ indicates the qualitative manner at which this saturation is achieved. These parameters allow for fine-tuning to specific experimental observations. 

Finally, a friction term employed in many biofilm multiphase models is
\begin{equation}
    \xi_B(\theta_s) = h\theta_s(1-\theta_s)\,.
\end{equation}
This choice has become popular in the literature \citep{byrne2003modelling,cogan2004role,cogan2005channel,cogan2010multiphase, sorribes2019biomechanical}, and is often justified by the idea that friction should vanish if either phase, $\theta_s$ or $\theta_m$, is absent. While it is an intuitive notion, we will show in section \ref{sec:poiseuille} that this friction coefficient does {\em not} produce the physically realistic behavior of no-slip velocity on fully developed solid surfaces. To produce this behavior, it is necessary that friction {\em dominates}, not vanishes, in the limit $\theta_m \to 1$.

\begin{figure}[h]
    \centering
    \includegraphics[scale=0.8]{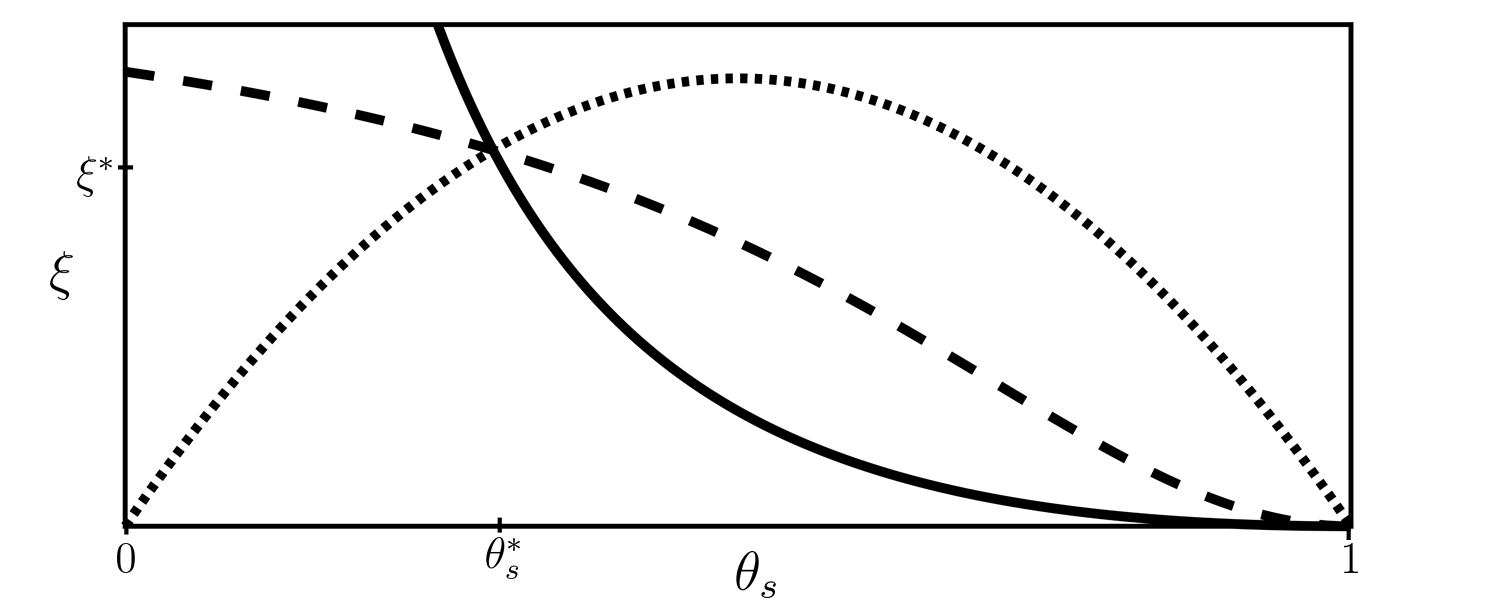}
    \caption{Comparison of friction terms $\xi$. $\xi_{KC}$ (solid) is singular in the limit $\theta_s\to 0$, $\xi_H$ (dash) is non-singular in the porous limit ($K=0.5$, $n=2$) and $\xi_B$ (dot) yields maximum friction when both phases are present in equal amounts. All terms have been normalized by choosing $h$ such that $\xi(\theta_S^*)=\xi^*$.}
    \label{fig:frictioncompare}
\end{figure}

A visual comparison of these three friction coefficients is shown in figure \ref{fig:frictioncompare}. For the sake of comparison, we have chosen the constant $h$ so that the three curves intersect at a reference porosity $\theta_s^*$, i.e.~$\xi(\theta_s^*) = \xi^*$ in each case.
For $\xi_{KC}$ and $\xi_H$, the value $\theta_s^*$ can be interpreted as the percolation threshold, i.e.~the critical porosity below which the medium essentially behaves as impermeable to flow \citep{Golden1997}. 

\subsection{Model Summary} \label{sec:modelsummary}
The governing equations are now summarized for the reader. Rearranging equations \eqref{eq:inhomA}--\eqref{eq:inhomC} and applying incompressibility of $\vq_s = \theta_s \vv_s$ gives the following ADR evolution equations for aqueous chemical concentration:

\begin{align}
&\parD{\psi_A}{t}  = \overbrace{\frac{1}{\theta_s}\nabla\cdot\big(\kappa_A\nabla\psi_A\big)}^{\text{\small diffusion}} - \overbrace{\nabla\psi_A\cdot\vv_s}^{\text{\small advection}} - \overbrace{a r \psi_A\psi_B}^{\substack{\text{\small aqueous} \\ \text{\small reaction}}} - \overbrace{\psi_A\dot{\theta}_s/\theta_s}^{\substack{\text{\small precipitate} \\ \text{\small reaction}}} \label{eq:chemAsum} \\
&\parD{\psi_B}{t}  = \frac{1}{\theta_s}\nabla\cdot\big(\kappa_B\nabla\psi_B\big) - \nabla\psi_B\cdot\vv_s - b r \psi_A\psi_B - \psi_B\dot{\theta}_s/\theta_s \label{eq:chemBsum} \\
&\parD{\psi_C}{t}  =\frac{1}{\theta_s}\nabla\cdot\big(\kappa_C\nabla\psi_C\big) - \nabla\psi_C\cdot\vv_s + c r \psi_A\psi_B - (\psi_C-\alpha)\dot{\theta}_s/\theta_s \label{eq:chemCsum}
\end{align}
where $\alpha=(\rho_m-\rho_s)/M_C$. Meanwhile, the evolution equations for the fluid and solid phases can be summarized as
\begin{align}
& \theta_s  + \theta_m = 1\,, \label{eq:consofvolume} \\
& \rho_m \parD{\theta_m}{t} = R_m 
= \beta \psi_C\theta_s\mathcal{H}(\psi_C - \psi_C^*) \, .
\label{eq:massbalanceMsum}
\end{align}
The momentum equation is a Brinkman equation with variable permeability
\begin{align}
    &\eta \nabla^2\vq_s - h\frac{\theta_m^2}{\theta_s^2}\vq_s = \theta_s \nabla P \label{eq:momentum} \\
    &\nabla\cdot\vq_s = 0 \label{eq:incompressQ}
\end{align}
where we have employed the $\xi_{KC}$ friction term. We will now consider these coupled PDEs in a simplified setting that will permit exact solutions.

\section{Analysis of Reduced Model} \label{sec:analysis}


Analysis of any complicated system is aided by reduction into a form that is analytically tractable. Inspired by microfluidic experiments, we assume that some chemostat controls the influx of reactants' molarity far upstream. All variables are kept constant along the longitudinal axis by neglecting diffusion and assuming parallel flow. This requirement on parallel flow also means that the reaction takes place everywhere along the longitudinal axis simultaneously. Finally, we assume that the precipitation threshold is negligible. 


Applying these assumptions to the governing equations reduces the system considerably so that it becomes a Poiseuille analysis; these assumptions generate the following reduced system
\begin{subequations} \label{eq:reducedsystem}
\begin{align}
&\dot{\psi}_C =  c r\psi_A\psi_B - (\psi_C - \alpha)\dot{\theta}_s/\theta_s  \label{eq:redC}\\
&\theta_s  + \theta_m = 1 \label{eq:redvol}\\
&\dot{\theta}_s = -\beta \psi_C \theta_s/\rho_m \label{eq:redthetas}\\
&\eta\parDD{q_y}{x} - h\frac{\theta_m^2}{\theta_s^2}q_y = \theta_s G\,. \label{eq:redmomentum}
\end{align}
\end{subequations}
where only $\psi_C$, $\theta_s$, $\theta_m$ and $q_y$ are unknown. Note that the longitudinal axis is chosen to be $y$ such that only this component of the Darcy velocity $\vq_s = q_x\mathbf{\hat{x}} + q_y\mathbf{\hat{y}}$ remains. In section \ref{sec:fixedpoint} we obtain an analytic estimate on the rate of membrane formation by treating $\psi_C$ and $\theta_m$ as a planar dynamical system. Then in section \ref{sec:poiseuille} we solve equations \eqref{eq:reducedsystem} with a combination of analytic and numerical methods to visualize how the membrane affects the flow profile in time.

\subsection{Fixed Point Analysis} \label{sec:fixedpoint}
Our approach is to linearize the reduced model system, then perform an eigenvalue analysis about a steady state fixed point. The benefit of this is the eigenvalue will correspond to the rate that membrane develops, a quantity that is possible to measure experimentally.

\begin{figure}[h]
	\centering
	\includegraphics[scale=0.9]{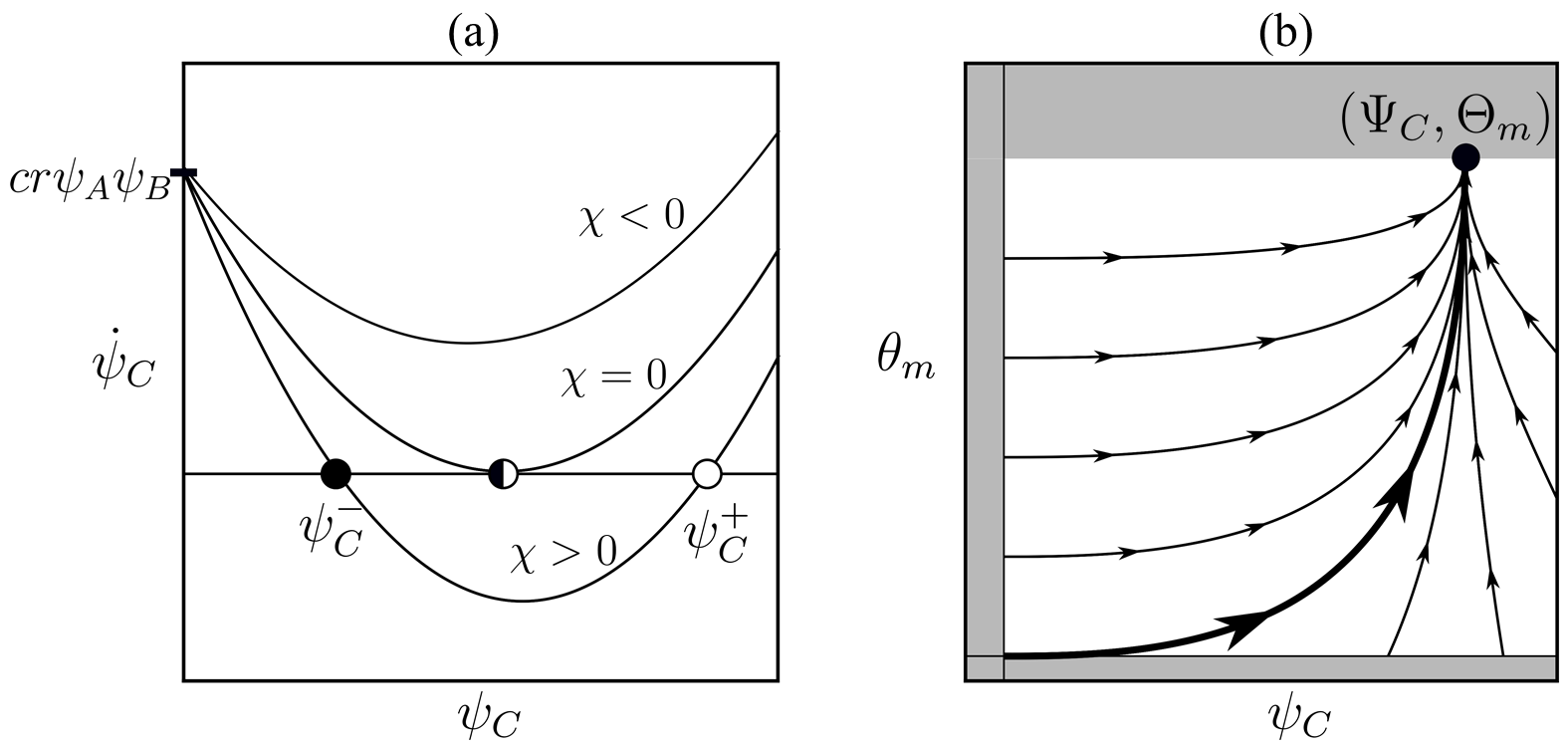}
	\caption{Dynamical system for $\psi_C$ and $\theta_m$. (a) Qualitative stability of $\dot{\psi}_C(\psi_C)$ ODE. The left equilibrium $\psi_C^-$ is stable and exists for $\chi\geq 0$. (b) Visualization of planar dynamical system. The thicker line corresponds to homogeneous initial conditions for $\psi_C$ and $\theta_m$. Shaded region is outside of the domain of $(\psi_C,\theta_m)\in[0,\infty)\times[0,1]$.}
	\label{fig:dynsysC}
\end{figure}

Before doing a fixed point analysis, it is helpful to understand the conditions on which the existence and stability of fixed points depends. To do so, eliminate the explicit dependence of $\dot{\psi}_C$ on volume fraction and replace all $\dot{\theta}_s$ in equation \eqref{eq:redC} with equation \eqref{eq:redthetas} to obtain a quadratic ODE of the form
\begin{equation}
\dot{\psi}_C = \frac{\beta}{\rho_m}\psi_C^2 - \frac{\alpha\beta}{\rho_m}\psi_C + cr\psi_A\psi_B\,. \label{eq:quadODE}
\end{equation}
which is an ODE in time alone, as the $x\text{-dependence}$ of $\psi_A$ and $\psi_B$ are determined by the initial conditions. Examining the qualitative behavior of this ODE by considering $\dot{\psi}_C=\dot{\psi}_C (\psi_C)$, it is quadratic in $\psi_C$, intercepts the $\dot{\psi}_C$ axis at $cr\psi_A\psi_B\geq 0$, is concave up, and has equilibria at 
\begin{equation}
\psi_C^{\pm} = \frac{1}{2}\left(\alpha \pm \frac{1}{\beta}\sqrt{\alpha^2\beta^2 - 4c r\rho_m \beta \psi_A\psi_B} \right)
\end{equation}
whose existence depends on the sign of 
\begin{equation}
\chi = \alpha^2\beta^2 - 4 c r \rho_m \beta\psi_A\psi_B\,. \label{eq:fixedptcond}
\end{equation}
If $\chi>0$, equation \eqref{eq:quadODE} will have two fixed points, for $\chi=0$ these fixed points coalesce, and for $\chi<0$ there are no fixed points and $\dot{\psi}_C$ will grow without bound; see figure \ref{fig:dynsysC}(a).

We now ask the question of whether fixed points exist in the reduced system, i.e.~$\chi\geq 0$ or $\alpha^2\beta \geq 4cr\rho_m\psi_A\psi_B$? We interpret this condition based on the physical meaning of the parameters: $\alpha=(\rho_m-\rho_s)/M_C$ has dimension of molarity and is $O(10 \,\text{M})$ where $\text{M}$ refers to molar units, $1 \,\text{M}=\text{1 mol}/\text{liter}$; for example, using the reaction system in the introduction gives $\alpha\approx 30 \, \text{M}$. We note that a similar analysis also justifies neglecting the precipitation threshold $\psi_C^*$, as $\psi_C^*\approx 0.001 \text{ M}$. Although both $r$ and $\beta$ scale the rates of the aqueous and precipitate reactions, respectively, they have different units. $r$ has units of volume per time while $\beta$ has units of mass per time. Because experimental values of $r$ and $\beta$ are expensive to acquire, for the sake of this simplified analysis we will assume that $\beta\approx r\rho_m$ such that their effects don't impact the sign $\chi$. The stoichiometric coefficient $c$ for $C\text{(aq)}$ can be assumed $O(1)$. Finally, examine the reactants $\psi_A$ and $\psi_B$. Most experiments in microfluidic chambers use molar concentrations with an upper bound of $O(1 \,\text{M})$; for example, in \citet{ding2016wavy} the maximum concentration of reactants was $0.5\, \text{M}$. Therefore, using parameter values taken from experiments, $\chi > 0$ and fixed points exist for the reduced system. 

We now consider the planar dynamical system in phase space $(\psi_C,\theta_m)\in[0,\infty)\times[0,1]$ with fixed points $(\Psi_C,\Theta_m)=(\psi_C^-,1)$. The dynamical system is
\begin{subequations} \label{eq:dynsystem}
\begin{align}
\dot{\psi}_C & = f(\psi_C,\theta_m) = \frac{\beta}{\rho_m}\psi_C^2 - \frac{\alpha\beta}{\rho_m}\psi_C + cr\psi_A\psi_B  \label{eq:psiF}\\
\dot{\theta}_m & = g(\psi_C,\theta_m) = \beta\psi_C(1-\theta_m)/\rho_m \label{eq:thetamG}
\end{align}
\end{subequations}
where $r$, $c$, $\rho_m$, $\alpha$, $\beta$, $\psi_A$, and $\psi_B$ are assumed to be known and constant. The eigenvalues of the Jacobian generated by equations \eqref{eq:dynsystem} evaluated at the fixed points provides information about the rate of growth of $\psi_C$ and $\theta_m$. This particular eigen-system is simple to interpret because the eigenvectors align with the coordinate axes and therefore the eigenvalues correspond to the rates that the physical variables $(\psi_C,\theta_m)$ approach their equilibria. These rates are given by:
\begin{equation}
\lambda_{\psi_C} = -\frac{1}{\rho_m}\sqrt{\chi}, \qquad \lambda_{\theta_m} = -\frac{1}{2}\left(\frac{\alpha\beta}{\rho_m}+\lambda_{\psi_C}\right)\, . \label{eq:eigen}
\end{equation}
Both eigenvalues are negative, and therefore the fixed point is stable, because \mbox{$\alpha\beta/\rho_m +\lambda_{\psi_C} = \alpha\beta/\rho_m - \sqrt{\chi}/\rho_m > 0$} always. For this same reason, it is true that $|\lambda_{\theta_m}| < |\lambda_{\psi_C}|$, meaning the membrane volume fraction approaches its fixed point at a slower rate than the aqueous product. This agrees with our intuition, as the conversion of $C\text{(aq)}$ to $C\text{(s)}$ means we would expect $\theta_m$ production to lag behind $\psi_C$.

\subsection{Poiseuille Analysis} \label{sec:poiseuille}
We now solve the equations to see how the solvent velocity transitions from one-channel to two-channel flow. An exact solution for $\psi_C(x,t)$ can be found by solving a Ricatti equation with constant coefficients. This solution $\psi_C$ can be integrated using elementary functions, so $\theta_s(x,t)$ can be found exactly because equation \ref{eq:redthetas} is separable; $\theta_m(x,t)$ can then be found using the no-void assumption. Finally, we solve the variable-coefficient BVP for $q_y$ numerically using finite differences.

For initial conditions, let $\psi_C(x,t) = 0$ and $\theta_s(x,t)=1$. Fix $\psi_A^0(x)$ and $\psi_B^0(x)$ to be piecewise-constant values in $x$ such that there is only a middle region in $x\in(0,L)$ in which the reactants $A$ and $B$ are simultaneously present. Given $\psi_A^0(x)$ and $\psi_B^0(x)$, the Ricatti equation \eqref{eq:redC} can be solved exactly for $\psi_C(x,t)$ (see appendix \ref{app:ricattisolve}): 
\begin{equation}
    \psi_C(x,t) = \gamma_1\gamma_2\frac{\rho_m}{\beta}\left(\frac{e^{\gamma_2 t} - e^{\gamma_1 t}}{\gamma_2e^{\gamma_1 t} - \gamma_1e^{\gamma_2 t}}\right) \label{eq:psicExact}
\end{equation}
where 
\begin{equation}
\gamma_{1,2}(x) = \frac{1}{2}\left(-\alpha\beta \pm \frac{\sqrt{\chi}}{\rho_m}\right)\, .
\end{equation}
Because both the antiderivative of $\psi_C$ can be given in terms of elementary functions (see appendix \ref{app:ricattisolve}) and equation \ref{eq:redthetas} is separable, we can also solve for solvent volume fraction $\theta_s(x,t)$ exactly:
\begin{equation}
\theta_s(x,t) = \frac{\gamma_1e^{\gamma_2 t} - \gamma_2 e^{\gamma_1 t}}{\gamma_1-\gamma_2}
\end{equation}
where have implemented the initial condition $\theta_s(x,0) = 1$. Then, the membrane volume fraction $\theta_m(x,t)$ can be computed easily using the no-void assumption.

Until this point, all solutions in space have been treated independently. The effect of variations in space is taken into account when solving for the longitudinal component of the Darcy velocity, $q_y$. We numerically solve the momentum equation \eqref{eq:redmomentum} for $q_y$ using a finite difference method. The $x$ domain is discretized into $N$ intervals of equal width $\Delta x=1/N$ such that $x_j=j\Delta x$, $j=1,\dots,N-1$, and use centered-difference approximations to the derivatives. Note that $q_y(x_0)=q_y(x_N)=0$ due to the no-slip boundary conditions. This discretization results in a tridiagonal linear system which can be solved in $O(N)$ complexity by using the Thomas algorithm \citep[see][pp. 78-79]{strikwerda2004finite}. The velocity is constrained to satisfy constant flux in accordance with experiments, which mathematically is represented by 
\begin{equation}
\int_0^Lq_y\,dx = \text{constant}\,. 
\end{equation}
This constant-flux condition allows the computation of $G(t_n)$ at each time step. The authors use the Julia programming language to solve the BVP \citep{bezanson2017julia}.

\begin{figure}[h]
    \centering
    \includegraphics[width= 0.9 \linewidth]{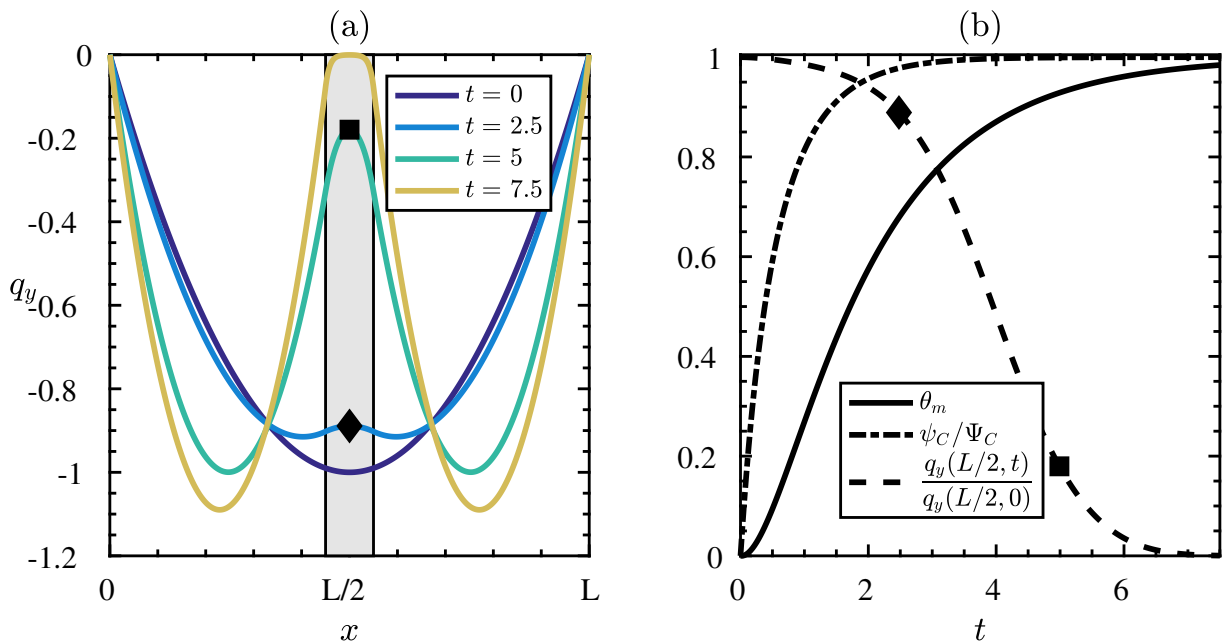}
    \caption{Developing membrane affects fluid flow. (a) Flow profile in a 1D channel transitions from one-channel to two-channel flow. The reaction region is shaded. (b) Relevant variables evaluated in the reaction region at $x=L/2$, normalized for legibility; $\psi_C$ develops first, followed by $\theta_m$, which when large enough triggers the transition from one- to two-channel flow. The percolation threshold $\theta_s^*$ is set to $\theta_s^*=0.3$, which is why negligible change in fluid velocity is seen until $\theta_m\approx 0.7$.}
    \label{fig:r1_plots}
\end{figure}

The developing membrane for the 1D reduced model geometry is shown in figure \ref{fig:r1_plots}(a). Membrane develops within the shaded region, which in this example is 10\% of the domain. Because the membrane has finite width, the constant-flux condition causes the pressure gradient $G(t)$ to increase with the developing membrane, causing the maximum speed for the two-channel flow to be slightly higher than the maximum speed for the one-channel flow. In this sense, the developing membrane splits the domain into two symmetric one-channel flows. 

Figure \ref{fig:r1_plots}(b) shows the three main variables of the reduced model as functions of time, evaluated in the middle of the reaction region ($x=L/2$). The $\psi_C$ variable increases immediately due to the presence of $\psi_A$ and $\psi_B$. The membrane initially has zero growth rate due to the absence of $\psi_C$, and grows at a slower rate than $\psi_C$. This ordering on the growth rates matches our expectations from the eigenvalue analysis of section \ref{sec:fixedpoint}. The $q_y$ curve demonstrates the transition from one-channel to two-channel pipe flow by measuring the normalized value in the middle of the pipe as a function of time. By comparing $q_y$ with $\theta_m$, one can see the effect of the percolation threshold $\theta_s^*=0.3$. After the solvent volume fraction declines past this value, the fluid velocity begins to react more strongly to precipitating membrane. 

\begin{figure}[h]
    \centering
    \includegraphics[width=\linewidth]{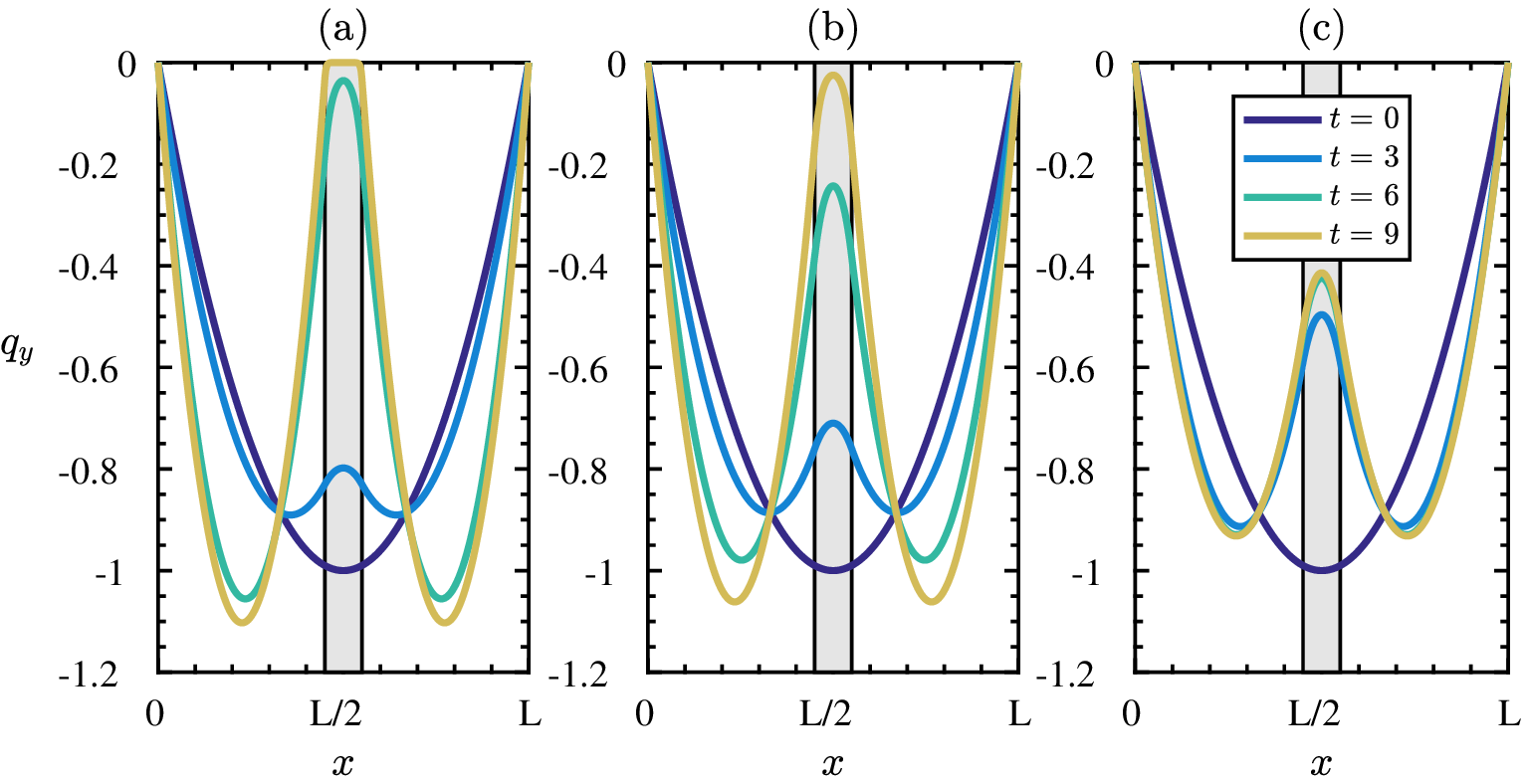}
    \caption{Effect of singular friction term. (a) Kozeny-Carman friction term, (b) Hill friction term and (c) generic friction term $\xi_B=h\theta_s\theta_m$ common in biofilm models. The coefficients $h$ are scaled so as to make the three coefficients comparable in strength. The first two friction terms produce the desired no-slip on the membrane interface and the third, while affecting the fluid flow, does not generate the desired no-slip boundary condition.}
    \label{fig:r2_plots}
\end{figure}

Figure \ref{fig:r2_plots} demonstrates the effect of using different friction coefficients. Both figures \ref{fig:r2_plots}(a,b) demonstrate the desired no-slip behavior in the membrane limit by using $\xi_{KC}$ and $\xi_H$, respectively. In figure \ref{fig:r2_plots}(c) the effect of the friction coefficient $\xi_B$ is shown. While there is some effect on the flow profile, $\xi_B$ does not demonstrate the no-slip condition on the membrane. While the $\xi_B$ term was developed primarily for high permeability applications, our framework was developed to capture the transition from purely fluid behavior, to partially permeable, to a fully-developed impermeable solid. As demonstrated, this full transition requires either the $\xi_{KC}$ or $\xi_H$ friction coefficient.

\begin{figure}[h]
    \centering
    \includegraphics[width=\linewidth]{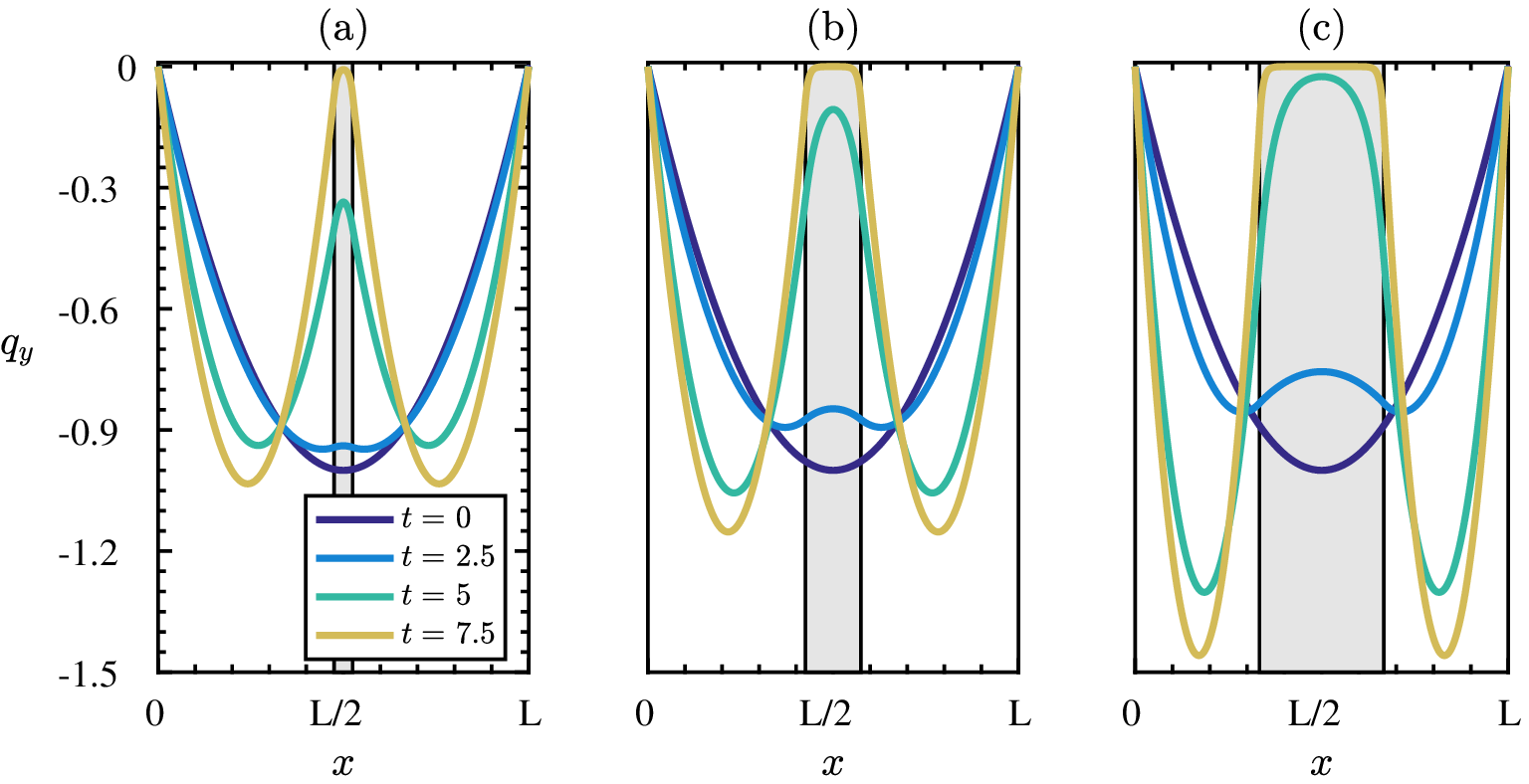}
    \caption{Effect of increasing membrane thickness. As a percentage of the domain length, membrane width is (a) $5\%$, (b) $15\%$, and (c) $33\%$. The increasing maximum flow speed is due to the constant-flux constraint, and is analogous to that what would occur if the membrane boundaries were prescribed \emph{a priori} in a single-phase flow.}
    \label{fig:r3_plots}
\end{figure}

Figure \ref{fig:r3_plots} shows three flows with reaction regions of various sizes, and therefore different width of membranes. The initial flow profiles of all are equivalent, as the reaction has not yet occurred and no membrane is present. However, as membrane develops, the constant-flux condition requires that for regions with thicker membranes, the flow velocity must increase in the non-reacting regions to compensate for the loss of flux in the membrane region. These results demonstrate that, once the membrane is fully developed, the flow domain treats the membrane portion as a no-slip boundary and the prescribed constant-flux conditions lead to the expected results from single-phase fluids.

\section{Conclusion} \label{sec:discussion}
Typically, no-slip conditions on boundaries must be prescribed \emph{a priori} when modeling fluid flows. However, motivated by microfluidic experiments of a precipitate reaction, there exist situations where solid materials develop dynamically and so the exact location of such a boundary is not known \emph{a priori}. This situation exposes a deficiency in modeling techniques that rely on knowledge of where no-slip boundaries exist.

To formulate the model, we assumed conservation of mass for the entire chemistry-fluid-membrane system. The reacting chemical species were assumed to exist in the fluid phase as volumeless scalar fields subject to mass flux determined by a combination of advection with the fluid flow and diffusion down gradients in molarity. The careful consideration of changing solvent volume inside the domain led to extra terms being included in the reaction equations, and in particular conservation of mass for the entire system was used to determine the closure for the aqueous product's reaction equation. A momentum equation for the fluid velocity followed the usual Cauchy stress formulation with the slight modification that all tensor quantities depend on the Darcy velocity $\vq_s=\theta_s\vv_s$, not simply the tracer velocity $\vv_s$ as is usual in the single- multi-phase fluids literature. This choice, paired with incompressibility of the fluid Darcy velocity, led to a simplification of the momentum balance to a variable-coefficient Brinkman equation.

In order to demonstrate that the model reflects the expected behaviors, we first sought growth rate estimates from a linearized form of reduced equations. Then, to show that the dynamically-generated no-slip boundary corresponded to development of the membrane, we performed what was essentially a Poisseuille analysis on the reduced model to show that the no-slip behavior agreed qualitatively with expected behaviors, specifically the recovery of one-to-two channel transitions and the effect of membrane width paired with a constant-flux condition.

We examined three potential terms to be used in the variable-coefficient Brinkman friction term, both by a direct comparison as functions of fluid volume fraction and by examining their effect on fluid flow from the perspective of how well they demonstrated no-slip behavior. These results demonstrated that, qualitatively, the friction terms derived from Karman-Cozeny relationship and using Hill functions gave no-slip flow behavior on the developed membrane. Additionally, the percolation threshold $\theta_s^*$ can be chosen to reflect specific permeability properties of the structure under investigation. Both of these friction coefficients were preferable to the term commonly employed in biofilm models because $\xi_B$ does not recover the no-slip behavior in low-permeability regions. 

To our knowledge, this is the first model that qualitatively captures the fluid-structure dynamics of a precipitate reaction in a low-Re environment where the dynamically-developing precipitate significantly affects the surrounding fluid flow. 
Future work will focus on numerical simulation of the full model in various geometries to be used as a predictive tool for experimentalists. In particular, exploring sufficient modeling conditions to generate the asymmetric growth in membrane in a 2D setting analogous to the experimental microfluidic domains was something that we were not able to explore in this paper, as diffusion was ignored to reduce the model equations. A model capable of accurately capturing microfluidic experiments would be valuable to researchers using these devices to study precipitate reactions at the microscale and ultimately useful in examining origin of life theories.

\section*{Acknowledgements}
P.S.E. is supported by the National Science Foundation (NSF) Graduate Research Fellowship under Grant 1449440. M.N.J.M. is supported by Simons Collaboration Grant 524259. N.G.C. is supported by NSF-CBET 1510743. Q.W. and O.S. are supported by NSF Grants 1609495 and 1565734. 

\appendix

\section{Solving Ricatti's Differential Equation} \label{app:ricattisolve}

The solution technique to Ricatti's differential equation \citep[see][p. 97]{tenenbaum1985ordinary} is not very well known, so the derivation is stated here for the interested reader. A homogeneous 1st order ODE is called a Ricatti equation if it is quadratic in the unknown, i.e.
\begin{equation}y'(x) = q_0(x) + q_1(x)y(x) + q_2(x)y^2(x)\,. \label{eq:genricatti} \end{equation}
If $q_0(x) \equiv 0$, this reduces to Bernoulli's equation \citep[see][pp. 95-96]{tenenbaum1985ordinary}. In general, one can transform Ricatti's equation to an equivalent 2nd order linear differential equation.  In this appendix, we detail this transformation and explicitly solve for the case of constant coefficients $q_0$, $q_1$ and $q_2$.

First, define a new variable $v$ by  
\begin{equation}v = q_2y \end{equation}
so that equation \eqref{eq:genricatti} becomes
\begin{equation}v' = v^2 + Rv + S \end{equation}
where $R = q_1 + q_2'/q_2$ and $S = q_0q_2$. Then, introduce another variable, $u$, related to $v$ via a Cole-Hopf transform:
\begin{equation}v = -\frac{u'}{u} \end{equation}
and now the original equation, in terms of $u$, becomes
\begin{equation}
u''-Ru'+Su = 0\,. \label{eq:ricattilinear}
\end{equation}
For constant coefficients $R$ and $S$, equation \eqref{eq:ricattilinear} can be solved exactly. Using terminology from the present paper, the constant coefficient Ricatti equation is
\begin{equation}
\dot{\psi}_C =  c r \psi_A\psi_B - \frac{\alpha\beta}{\rho_m} \psi_C + \frac{\beta}{\rho_m}\psi_C^2
\end{equation}
where $\psi_C=\psi_C(t)$ is the unknown variable and $r$, $c$, $\rho_m$, $\alpha$, $\beta$, $\psi_A$, and $\psi_B$ are fixed parameters. Relating this to equation \eqref{eq:genricatti}, let $y=\psi_C$, $q_0 = c r \psi_A\psi_B$, $q_1=-\alpha\beta/\rho_m$, and $q_2=\beta/\rho_m$. Therefore, in the final equation, 
$R = q_1 + q_2'/q_2 = -\alpha\beta/\rho_m$ and $S = q_0q_2 = c r \beta \psi_A\psi_B/\rho_m$. 

The solution to equation \eqref{eq:ricattilinear} when $R$ and $S$ are constant depends on the eigenvalues of its corresponding characteristic equation. More specifically, it depends on the sign of the determinant of the root of the characteristic equation $R^2-4S=\chi/\rho_m^2$, where $\chi$, defined in section \ref{sec:fixedpoint}, was determined to be positive for physically realistic parameter values.

The solution to this case -- where the characteristic equation to equation \eqref{eq:ricattilinear} has two real, distinct roots -- can be found in any introductory book on ODEs \citep[see][pp. 137 - 143]{boyce2008elementary}, but for completeness the solution is detailed here with homogeneous initial conditions $y(0) = 0$:
\begin{equation}y(t) = \frac{\gamma_1\gamma_2}{q_2}\left[\frac{e^{\gamma_2 t} - e^{\gamma_1 t}}{\gamma_2 e^{\gamma_1 t} - \gamma_1e^{\gamma_2 t}}\right]\,. \end{equation}
where 
\begin{align}
\gamma_{1} & = \frac{R + \sqrt{R^2-4S}}{2}, & \gamma_{2} & = \frac{R - \sqrt{R^2-4S}}{2} \\
R & = q_1, & S & = q_0q_2.
\end{align}
and the antiderivative of this solution is given by
\begin{equation}
\int y(t) \,\, dt = -\frac{1}{q_2}\log(\gamma_1 e^{\gamma_2 t} - \gamma_2 e^{\gamma_1 t}) + C
\end{equation}
where $C$ is an arbitrary constant of integration.
\bibliographystyle{plainnat}
\bibliography{membrane}

\end{document}